# Airbnb in tourist cities: comparing spatial patterns of hotels and peer-to-peer accommodation

JAVIER GUTIÉRREZ
Departamento de Geografía Humana.
Universidad Complutense de Madrid
javiergutierrez@ghis.ucm.es

JUAN CARLOS GARCÍA-PALOMARES
Departamento de Geografía Humana.
Universidad Complutense de Madrid
jcgarcia@ucm.es

GUSTAVO ROMANILLOS
Departamento de Geografía Humana.
Universidad Complutense de Madrid
gustavro@ucm.es

MARÍA HENAR SALAS-OLMEDO
Departamento de Geografía Humana.
Universidad Complutense de Madrid
msalas01@ucm.es

**Abstract:**

In recent years, what has become known as collaborative consumption has undergone rapid expansion through peer-to-peer (P2P) platforms. In the field of tourism, a particularly notable example is that of Airbnb, a service that puts travellers in contact with hosts for the purposes of renting accommodation, either rooms or entire homes/apartments. Although Airbnb may bring benefits to cities in that it increases tourist numbers, its concentration in certain areas of heritage cities can lead to serious conflict with the local population, as a result of rising rents and processes of gentrification. This article analyses the patterns of spatial distribution of Airbnb accommodation in Barcelona, one of Europe's major tourist cities, and compares them with the accommodation offered by hotels and the places most visited by tourists. The study makes use of new sources of geolocated Big Data, such as Airbnb listings and geolocated photographs on Panoramio. Analysis of bivariate spatial autocorrelation reveals a close spatial relationship between the accommodation offered by Airbnb and the one offered by hotels, with a marked centre-periphery pattern, although Airbnb predominates over hotels around the city's main hotel axis and hotels predominate over Airbnb in some peripheral areas of the city. Another interesting finding is that Airbnb capitalises more on the advantages of proximity to the city's main tourist attractions than does the hotel sector. Finally, it was possible to detect those parts of the city that have seen the greatest increase in pressure from tourism related to Airbnb's recent expansion.

1. Introduction

The last few years have seen the emergence of the so-called sharing economy (also known as collaborative consumption), within the framework of a lifestyle in which more importance is attached to sharing goods than to owning them ("using rather than owning"). With this system, consumers benefit from lower costs for using goods and services at the same time as they avoid wasting resources (Leismann et al., 2013). Collaborative consumption has been driven by the development of Internet platforms that facilitate peer-to-peer relations. The Internet and especially Web 2.0 has brought about many new ways of sharing as well as facilitating older forms of sharing on a larger scale (Belk, 2014; Botsman y Rogers, 2011). Collaborative consumption could therefore be broadly defined nowadays as peer-to-peer-based activity for obtaining, giving, or sharing the access to goods and services, coordinated through community-based online services (Hamari et al., 2015).

One of the fields in which collaborative consumption has burst onto the scene with greater intensity is that of tourism, both in the travel sector (car-sharing) and that of accommodation (home exchanges and room/apartment rentals), the best-known platforms being BlaBlaCar and Airbnb, respectively. The exchange of accommodations between private individuals has historically developed informally, but the Internet, and more specifically Web 2.0, has allowed it to grow exponentially and take on new characteristics (Russo and Quaglieri, 2014). Peer-to-peer platforms in the field of accommodation go well beyond marketing and advertising the properties. They screen both parties, have access to the owners' inventories, manage rental bookings, collect payments and provide some form of insurance coverage for damages caused by the renters (Pizam, 2014). Compared to business models that connect the business with the consumer (B2C), such as Expedia or Booking.com for hotel bookings, the business model for these alternative platforms is based on direct contact between individuals (person to person or P2P), which may involve hosts and travellers (Airbnb) or people who wish to exchange their accommodation free of charge (Couchsurfing). Renters can obtain accommodations at lower prices from Airbnb than from hotels in most cities

Airbnb is the most successful P2P platform in the field of accommodation and defines itself on its website (www.airbnb.com) as "a trusted community marketplace for people to list, discover, and book unique accommodations around the world — online or from a mobile phone or tablet". It connects people who have space to spare (hosts) with those who are looking for a place to stay (guests). Airbnb reaches more than 2,000,000 listings in 190 countries, mainly entire apartments and homes (57%) and private rooms (41%). The Airbnb offering has grown exponentially. According to the company's own data, the number of listings rose from 50,000 in late 2010 to 200,000 by mid-2012, 300,000 by early 2013, more than one million worldwide by the end of 2014 and over two million by the end of 2015. Most of these properties are located in Europe and the USA. Airbnb's headquarter is located in San Francisco, where the company was founded. As a result of its rapid worldwide expansion, international offices have been progressively opened in some of the world's major cities: London, Paris, Berlin, Milan, Barcelona, Copenhagen, Dublin, Moscow, São Paulo, Sydney and Singapore. Although Airbnb remains privately held, its valuation of over $10 billion now exceeds that of well-established global hotel chains like Hyatt (Zervas et al., 2014). Airbnb revenues come from charging a fee to guests (between 10% and 12%) and hosts (3%).

As a disruptive innovation in the field of tourism accommodation[1,] Airbnb proposed a novel business model, built around modern internet technologies and Airbnb's distinct appeal, centred on cost-savings, household amenities and the potential for more authentic local experiences (Guttentag, 2013). Taking part in Airbnb has been defined as significantly different to 'mainstream' consumption in terms of meaningful life enrichment, human contact, access and authenticity. Experiencing a city and living like a local are aspects that are valued and sought after by Airbnb users (Yannopoulou et al., 2013). Most importantly, Airbnb's relatively low costs appear to be a major draw. Airbnb hosts are able to price their spaces very competitively because the hosts' primary fixed costs (e.g. rent and electricity) are already covered; hosts generally have minimal or no labour costs, do not usually depend solely on their Airbnb revenue and generally do not charge taxes (Guttentag, 2013).

Airbnb is at a clear disadvantage for competing with accommodation offered by the formal economy in terms of quality of service, staff professionalism, brand reputation and security for both hosts and guests. In order to offset some of these disadvantages Airbnb provides various services, such as a 24-hour customer telephone service and the publication of guests' reviews and ratings in order to improve the quality of service and create trust among users. Trust is also fostered through communication between hosts and guests by direct messaging and through users' profiles, which may display a photograph and include descriptive personal information (Guttentag, 2013). Nevertheless, these practices may determine the trustworthy perceived by hosts (Ert, Fleischer, & Magen, 2016) and also facilitate discrimination based on the seller's race, gender, age, or other aspects of appearance (Edelman and Luca, 2014). In the wake of acts of vandalism in some apartments, which have incurred heavy losses for the hosts, Airbnb has reacted by contracting an insurance policy (Host Guarantee) that provides protection for up to $1,000,000 worth of damage (www.airbnb.com).

The potential impacts of Airbnb on local economies are complex and difficult to measure. The results of the study by Fang et al. (2006) suggest that the entry of sharing economy benefits the entire tourism industry by generating new job positions as more tourists would come due to the lower accommodation cost. However, since low-end hotels are being shocked and replaced by Airbnb, the marginal effect decreases as the size of sharing economy increases. Zervas et al. (2014) have studied the impact of Airbnb's entry into the Texas market on hotel room revenue, estimating that in Austin, where Airbnb supply is highest, the impact on revenue is roughly 8-10%. This impact is non-uniformly distributed, with lower-priced hotels, and hotels not catering to business travel being the most affected segments. These results are consistent with the typical Airbnb client profile, that is, young, adventurous and budget-conscious tourists, who are familiar with the use of the Internet (Guttentag, 2013; Russo and Quaglieri, 2014). According to the estimations of Zervas et al. (2014), in Texas each additional 10% increase in the size of the Airbnb market resulted in a 0.37% decrease in hotel room revenue. On the other hand, in countries where Airbnb is less well-established, such as Korea, its effects on hotel revenue have so far been negligible (Choi et al., 2015).

---

[1] The disruptive innovation theory describes how products that lack in traditionally favoured attributes but offer alternative benefits can, over time, transform a market and capture mainstream consumers (Guttentag, 2013).

From the perspective of the spatial distribution of the Airbnb impacts within the cities, it has been argued that Airbnb listings are more scattered than hotels, so Airbnb guests may be especially likely to disperse their spending in neighbourhoods that do not typically receive many tourists (see Guttentag, 2014). Economic impact studies carried out by Airbnb show that most Airbnb properties (74%) are outside the main hotel districts, 79% of travellers want to explore a specific neighbourhood, and 42% of guest spending is in the neighbourhoods where travellers stayed (https://www.airbnb.co.uk/economic-impact).

Nevertheless, this possible dispersion may be compatible with a particular concentration of listings in the central areas of the cities, including areas not covered by hotels. As Zervas et al. (2014) point out, Airbnb can potentially expand supply wherever houses and apartment buildings already exist, in contrast to hotels, which must be built at locations in accordance with local zoning requirements. Therefore, expanding in historic centres would be easier for Airbnb than for hotels, which not only requires whole buildings to be available but also the relevant permits from the authorities. It is precisely these central areas of the city that register the greatest tourist concentrations, as confirmed by analyses carried out from geolocated photographs taken by tourists (García-Palomares et al., 2015). If Airbnb shows a clear tendency towards expansion in historic centres, then this could aggravate the problems of crowding[2] and tourism gentrification[3] that some of these areas have to support in certain heritage cities (Russo, 2002; Neuts and Nijkamp, 2012). This is not only a question of coexisting with ever-growing mass tourism, but homes would also suffer from rising rents brought about by the progressive expansion of Airbnb and other apartment rental platforms. The problem would not so much derive (at least for the time being) from the quantity of accommodations supplied by Airbnb in each city, but from its concentration in areas that are already subjected to strong pressure from tourism and the associated processes of gentrification.

The sharing economy has transformed many aspects of the tourism sector. Nevertheless, academic studies on Airbnb and its effects on the traditional tourist sector and cities are particularly scant. Guttentag (2013) studied Airbnb as a disruptive innovation in the accommodation sector. Zebras et al. (2014) and Choi et al. (2015) focused their attention on competition from Airbnb with the traditional accommodation sector. Yannopoulu et al. (2013) analysed the construction of user-generated brands (UGBs), using discursive and visual analysis of UGBs' social media material, taking Airbnb and CouchSurfing as examples. None of these studies examined the spatial distribution patterns of Airbnb listings. The present study attempts to identify the density of Airbnb listings in Barcelona, one of Europe's top tourist cities, and compare this with the density of hotel accommodation on offer. These results are subsequently correlated with the distribution of the main sightseeing spots in the city, identified from geolocated photographs from the Panoramio data source (see García-Palomares et al., 2015). GIS tools and spatial statistics were used to carry out the analysis, which specifically examines univariate and bivariate spatial autocorrelation.

---

[2] Crowding is linked with the concepts of carrying capacity and sustainability. According to The World Tourism Organization, carrying capacity is the maximum number of people that may visit a tourism destination at the same time without causing destruction of the physical, economic and socio-cultural environment and an unacceptable decrease in the quality of visitors' satisfaction. Carrying capacity is essentially a threshold that indicates the point at which a situation becomes unsustainable. Crowding is thus specifically seen as the violation of the sociocultural carrying capacity (Neuts and Nijkamp, 2012).
[3] Tourism gentrification has been defined as the transformation of a middle-class neighbourhood into a relatively affluent and exclusive enclave marked by a proliferation of corporate entertainment and tourism venues (Gotham, 2005). In this study the term tourism gentrification has a wider meaning, referring to the transformations in residential areas brought about by tourism without necessarily resulting in the formation of affluent and exclusive enclaves.

The article is structured as follows. After the introduction, Section 2 outlines the area of study, Sections 3 and 4 describe the data and the methodology, respectively, Section 5 shows the main results and Section 6 presents the conclusions and final remarks.

## 2. Study Area: Barcelona

The area selected for study was the city of Barcelona. Barcelona is a historic city that suffers serious problems from mass tourism and has a large concentration of accommodation on offer on the Airbnb website. Delimitation of the study area was based on the administrative reference unit: the municipality of Barcelona (Figure 1). This is a relatively compact area of 10,130 hectares around the traditional city. With 1,604,000 inhabitants, the population density of the municipality of Barcelona is high, with more than 158.3 inhabitants per hectare.

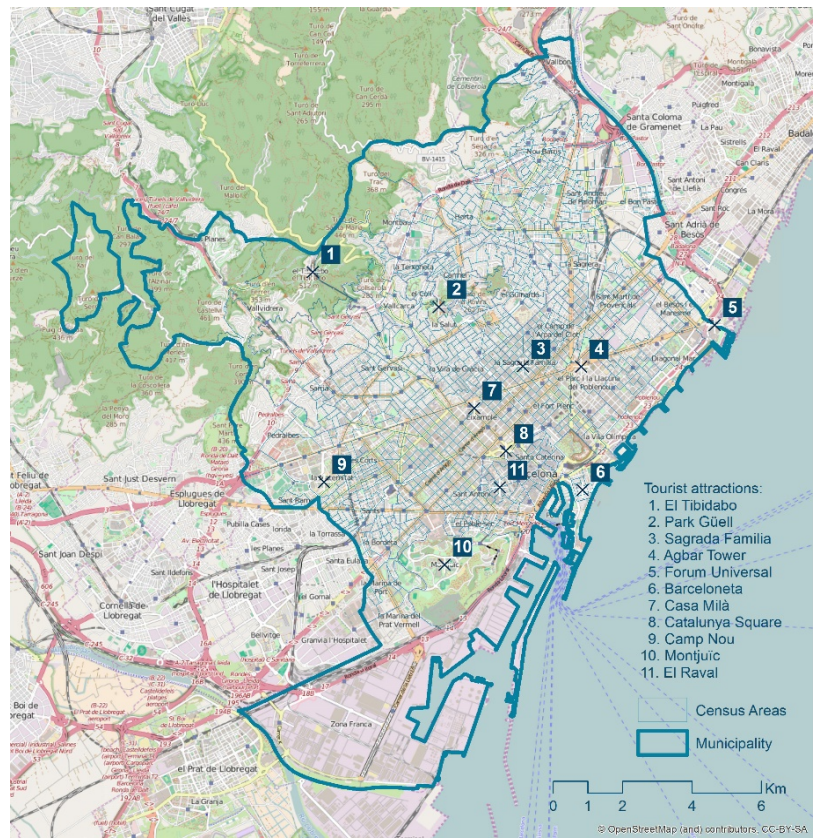

Figure 1: Municipality of Barcelona

In 2014, Barcelona was the fifth city in Europe in terms of the number of international tourists, behind only London, Paris, Berlin and Rome (European Cities Marketing, 2015[4]). On a global scale, it is among the twenty-five favourite city destinations for international tourism (Top Cities Destination Ranking 2013, Euromonitor International). Its popularity rose considerably after it hosted the 1992 Olympic Games. In 1990 the number of overnight stays totalled 3.8 million, involving 1.7 million tourists. In 2000, there were 7.9 million overnight stays and 3.1 million tourists. By 2014 this total had reached almost 17 million overnight stays, with 7.8 million tourists (79.5% of them international tourists). This huge influx of visitors has an enormous economic and social impact on the city, generating more than 26 million euros a day and more than 120,000 jobs in tourism (Barcelona Turisme: Barcelona Tourism Annual Report, 2014[5]), but also producing a high pressure on the city centre that led to a significant gentrification process (Cócola, 2015).

---

[4] http://www.cvent.com/events/ecm-benchmarking-report-2013-2014/event-summary-9dc7593d7ca947d995cd7ca658a0777d.aspx
[5] http://professional.barcelonaturisme.com/imgfiles/estad/Est2014b.pdf

## 3. Data

This research is mainly based on the analysis of the Airbnb geolocated data obtained from the Inside Airbnb website http://insideairbnb.com/. Inside Airbnb is an independent initiative and the data made available through its website are "*not associated with or endorsed by Airbnb or any of Airbnb's competitors*". The data utilises public information compiled from the Airbnb web-site, including not only the location of all Airbnb accommodations but also the availability calendar for 365 days in the future, and the reviews for each listing [6]. These data are available for more than 30 cities, including the main cities in Europe (London, Paris, Berlin, Madrid, etc.), the United States and Canada (New York City, San Francisco, Los Angeles, Washington D.C., Montreal, Vancouver, Toronto, etc.) and Australia (Sydney and Melbourne).

Data compiled for Barcelona refers to October 2015. Two files called *listings.csv* were downloaded. These contained ample information on each listing with respect to *Room Type* (entire homes/apartments; private vs shared rooms; number of bedrooms and beds), *Activity* (estimated nights/year; reviews/listings/month; reviews; estimated occupancy; price/night; estimated income/month), *Availability*, *Listings per Host*, etc. From the x, y coordinates stored in each record a point layer map was created in a geodatabase in ArcGIS with the location and features of each accommodation (Figure 2a). The map shows a clear concentration of points in the city centre. This spatial pattern is the result of an explosive growth, as illustrated in the supplementary video on the evolution of Airbnb accommodations in Barcelona (see video). By using the number of reviews per month as a proxy for the level of occupation of the accommodations, an average value of 1.48 was obtained for the centre of the city [7] and 1.14 for the rest, data that also expresses the higher pressure exerted by Airbnb on the city centre. With regard to the type of accommodation, 54% were entire homes/apartments, 45% were private rooms and only 1% were shared rooms, with prices averaging about 35 euros/bed. The database also reveals that this platform is not only used by private individuals but also by professionals. The proportion of Airbnb hosts who rent out more than one room or apartment is about 27% and 22% of the rooms or apartments are rented out by hosts who offer more than 5 accommodations.

Accommodation offered by Airbnb was compared with that of the city's hotels (Figure 2a). Data on hotels were taken from the Catalonia Tourism Registry[8], compiled and updated weekly by the regional government (*Generalitat de Catalunya*). The records for each hotel contain data on the number of rooms and beds available as well as the corresponding postal address. Geolocation of these data was carried out using ArcGIS address matching tools. There are 670 hotels in Barcelona offering more than 70,000 places. In the case of Airbnb, the number of accommodations is 14,500, with an offering of approximately 51,000 places (Table 1). However, the Airbnb data should be put into context. While the hotel offering covers 365 days of the year, Airbnb beds are available over less time. In Barcelona, the average availability of the listings is for 280 days a year (Table 2). Of these listings, 1,706 (12%) are available for fewer than 90 days a year.

In order to identify sightseeing hot spots, geolocated photographs from the Panoramio data source were used. The data were downloaded through the Panoramio website API[9] to obtain samples of all the photographs stored, and contained information about the geographic coordinates, the ID of the owner of the photograph, a url link to the photograph and the date on which it was uploaded. Downloading generated ".csv" files, which contained the geographical coordinates of the location of each of the photographs. These coordinates were used to create a layer of points for each of the locations in a GIS. Geolocated photographs were differentiated according to whether they had been taken by tourists or residents. We used the same criterion as Fischer for his Geotaggers' World Atlas and García-Palomares et al. (2015): if this period exceeded one month, then the photographs were attributed to residents; if the period was less than one month, then they were attributed to tourists. The number of photographs taken in Barcelona was more than 92,000, of which 28.5% were taken by tourists (Table 3). The spatial distribution of photographs taken by tourists is much more concentrated than that of those taken by residents, reflecting the location of the city's main tourist attractions (Figure 2b).

---

[6] All information on this source can be consulted at http://insideairbnb.com/about.html
[7] Ciutat Vella and Eixample districts.
[8]http://empresaiocupacio.gencat.cat/es/treb_ambits_actuacio/emo_turisme/emo_empreses_establiments_turistics/emo_registre_turisme_catalunya/emo_llistat_establiments_turistics/index.html
[9] http://www.panoramio.com/api/data/api.html

Table 1. Data on tourist accommodations in Barcelona: hotels versus Airbnb

|         | Hotels / Airbnb listings | Bedrooms | Beds   | Places |
|---------|--------------------------|----------|--------|--------|
| Hotels  | 670                      | 37,405   | 70,465 | 73,158 |
| Airbnb  | 14,539                   | 22,059   | 33,167 | 50,969 |

Source: InsideAirbnb and Generalitat de Catalunya.

Table 2. Basic data on listings offered by Airbnb in Barcelona

|                |      | Room type       |              |             | Total  |
|----------------|------|-----------------|--------------|-------------|--------|
|                |      | Entire home/apt | Private room | Shared room |        |
| Listings       |      | 7,816           | 6,566        | 157         | 14,539 |
| Price          | Mean | 111.8           | 39.3         | 27.3        | 78.2   |
|                | SD   | 140.2           | 26.8         | 21.5        | 110.5  |
| Availability   | Mean | 276.1           | 285.8        | 305.8       | 280.1  |
|                | SD   | 103.8           | 111.0        | 107.0       | 107.3  |
| Beds           | Mean | 3.1             | 1.3          | 3.6         | 2.3    |
|                | SD   | 2.0             | 1.0          | 3.4         | 1.8    |
| Reviews/listing/month | Mean | 1.28     | 1.41         | 1.18        | 1.34   |
|                | SD   | 1.31            | 1.56         | 1.51        | 1.43   |

Source: InsideAirbnb.

Table 3: Photograph statistics (Photographs/hectare)

|                       | Tourists' photographs | | Locals' photographs | | All photographs | |
|-----------------------|-----------------------|----------|---------------------|----------|-----------------|----------|
|                       | Total                 | Density* | Total               | Density* | Total           | Density* |
| Barcelona municipality| 26,361                | 1.50     | 66,114              | 3.75     | 92,475          | 5.25     |

Finally, population data from sections of the Barcelona municipal census were used. The population data were obtained from the official registry of inhabitants (Padrón del Instituto Nacional de Estadística: http://www.ine.es/ ), taking 2013 as the date of analysis.

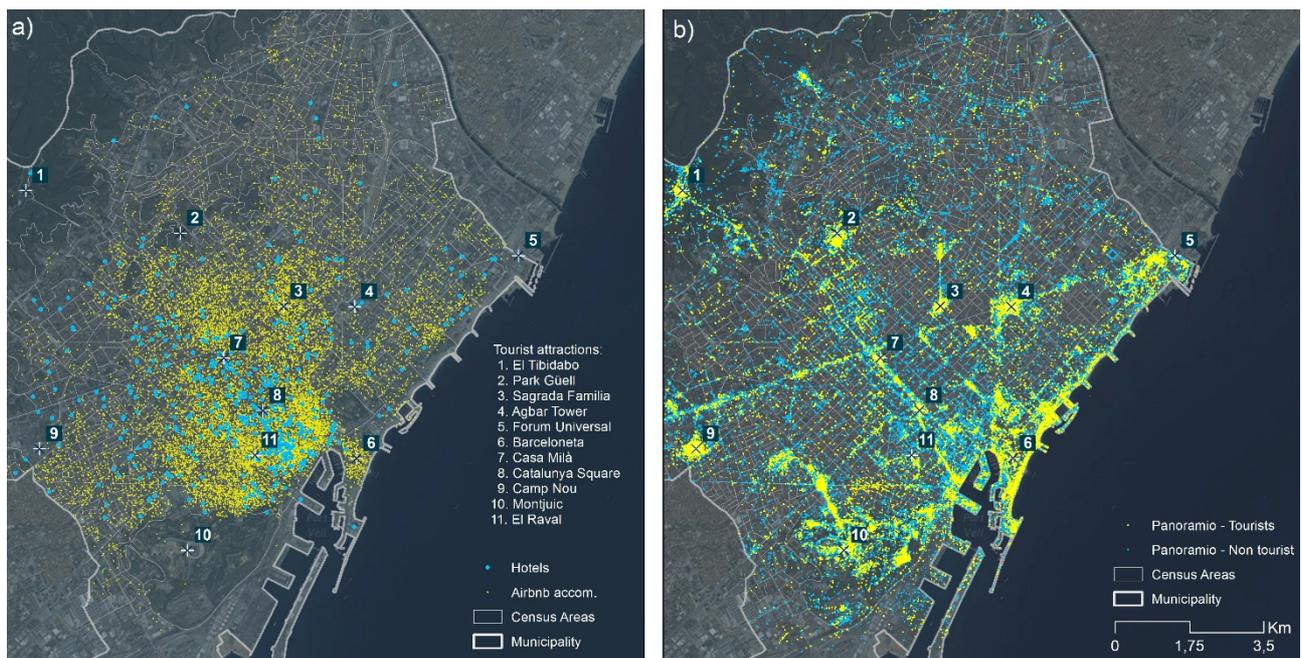

Figure 2: Location of hotel and Airbnb offers (a) and density of photographs taken by tourists and residents (b)

## 4. Methodology

The following methodology was employed to analyse the spatial distribution of Airbnb and Hotel accommodations and photographs:

a) Data aggregated by census tract were used to produce density maps and descriptive statistics, and thus determine the intensity and degree of concentration of the accommodations (differentiated by hotels and Airbnb) and tourist photographs.
b) Distribution data on hotel and Airbnb accommodations was normalised in order to eliminate the effect of different ranges in the variables so that distributions could be compared with each other.
c) Univariate spatial autocorrelation tools were used to identify the location and extent of spatial clusters of types of accommodations and tourist photographs.
d) Bivariate spatial autocorrelation tools were used to analyse spatial autocorrelation between accommodation types (hotels and Airbnb) and tourist photographs.
e) Finally, the rates of hotel and Airbnb accommodations per 1,000 inhabitants were obtained in order to analyse pressure from tourism on the resident population.

Aggregated data at census section level and density mapping are an initial visual approach to distribution of the accommodation offering, allowing the spatial patterns between Airbnb and hotel accommodation to be compared. The descriptive statistics enable measurements to be taken to determine the degree of concentration or dispersion of the types of accommodations. Data normalisation was used in order to map the degree of relative predominance of one type of offering over another in each of the census sections. To do this, the normalised densities of the Airbnb offering were deducted from the normalised densities of the hotels offered in each section.

Based on aggregated data by census sections, the location patterns were then analysed using spatial statistical indicators. Global Moran's I statistic was calculated to measure spatial autocorrelation based on feature locations and attribute values. Anselin Local Moran's I (LISA statistic) was used to identify local tendencies in the location of the different types of accommodation. With LISA analysis it was possible to distinguish High-High clusters (a high value surrounded primarily by high values), Low-Low clusters (a low value surrounded primarily by low values), and spatial outliers, either High-Low (high values surrounded primarily by low values) or Low-High (low values surrounded primarily by high values) (Anselin, 1995). Global and Local Bivariate Moran's I were used in order to measure spatial autocorrelation between variables and to identify spatial clusters in which the high values of one variable were surrounded by high values of the second (i.e. lagged) variable (high-high clusters) and so on.

Selection of the spatial interaction method requires special attention when computing spatial autocorrelation statistics. In this case we chose to consider the spatial interaction between observations within a 1 kilometre radius, that is, a typical 15-minute walk. Any observation within this radius would therefore be considered in the analysis with a weight inversely proportional to the distance separating the origin and the destination.

GeoDa was used to compute both univariate and bivariate Global and Anselin Local Moran's I. GeoDa is an open source package developed in the GeoDa Center for Geospatial Analysis and Computation [10]. This software provides all the necessary tools for performing spatial analysis in a simple way, even for non-experts or academics, at the same time as it generates high value and easy to understand outputs. As with other GIS software, GeoDa offers specific tools for univariate exploratory analysis. The main advantage is that it also allows the computation of a bivariate spatial Moran's I autocorrelation index (Anselin et al., 2006).

## 5. Results

### 5.1. Distribution of tourist accommodations

Maps and descriptive statistics of the distribution of accommodations by census sections are shown in Figure 3 and Table 4. The average number of Airbnb accommodations by census sections is 48, compared with 69 for hotels, with maximum values of around 600 accommodations for Airbnb and more than 2000 for hotels in some sections of the centre. The number of census sections with more than 200 accommodations is also much greater in the case of hotels (Table 4). The hotels are highly concentrated in the census sections comprising the Ramblas-Paseo de Gracia axis, certain areas dedicated to business and finance, like the Diagonal main street, or the coastal axis from the Barceloneta beach to the Forum. Outside these areas the availability is much lower. The differences in the distribution of Airbnb accommodations are not so marked, as shown by the coefficient of variation, which has much lower values for Airbnb than for hotels.

---

[10] GeoDa Center & Affiliated Software: https://geodacenter.asu.edu/software. Last visited on 03 February 2016.

Table 4: Statistics on the distribution of accommodation offered by hotels and Airbnb according to census areas.

|  | Airbnb | | | Hotels | | |
| --- | --- | --- | --- | --- | --- | --- |
|  | Total | Accommodations | Accommodations /ha | Total | Accommodations | Accommodations /ha |
| Count: | 1061 | 1061 | 1061 | 1061 | 1061 | 1061 |
| Minimum: | 0 | 0 | 0 | 0 | 0 | 0 |
| Maximum: | 175 | 616 | 64.0 | 48 | 2368 | 255.4 |
| Sum: | 14515 | 50969 | 6586.9 | 712 | 73158 | 6089.5 |
| Mean: | 13.7 | 48.0 | 6.2 | 0.7 | 69.0 | 5.7 |
| Standard Deviation: | 21.3 | 76.7 | 8.8 | 3.0 | 233.5 | 18.5 |
| CV: | 155.6 | 159.7 | 141.9 | 443.8 | 338.7 | 322.8 |
| Nº Census sections > 100 Accommodations | | | 153 (14.4%) | | | 153 (14.4%) |
| Nº Census sections > 200 Accommodations | | | 50 (4.7%) | | | 90 (8.5%) |

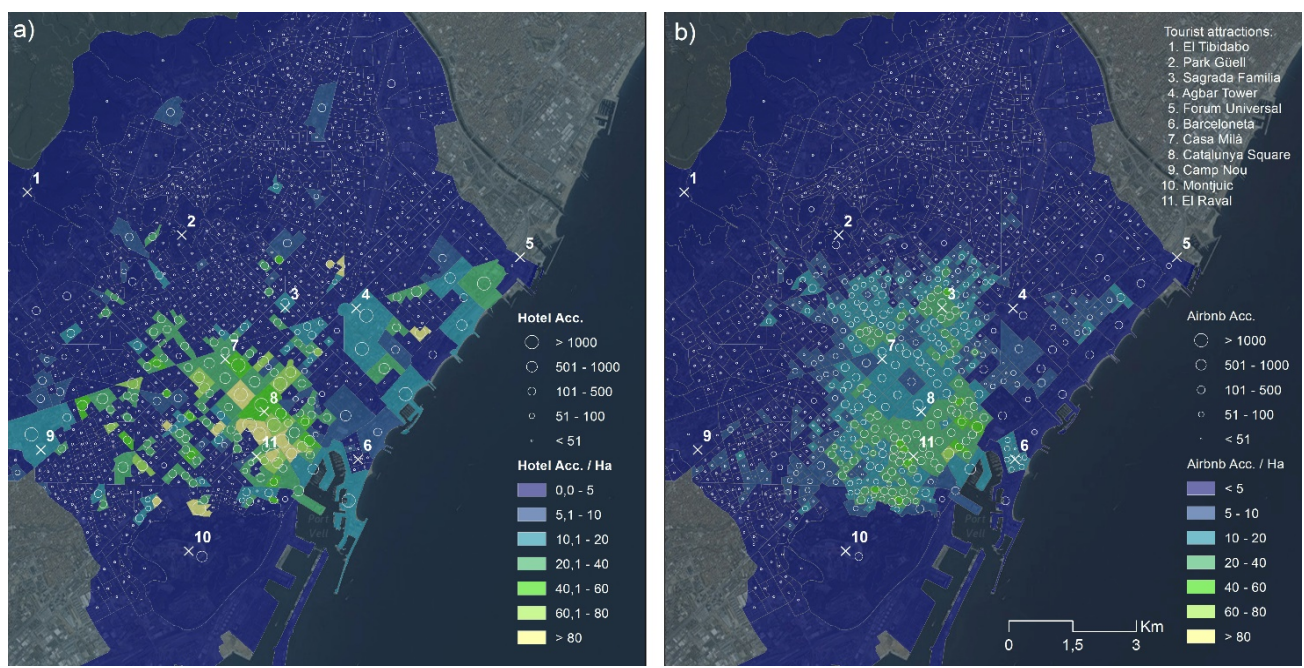

Figure 3: Total number and density of places, according to census sections: a) hotels, and b) Airbnb

In order to compare the distribution of hotel and Airbnb accommodation by census sections, Figures 4a and 4b show the density of accommodation places with normalised data. This cancels out differences in the ranges of the two variables, thus making them comparable. In contrast to the marked concentration of hotels on the Ramblas-Paseo de Gracia axis, Airbnb accommodation is found in a concentric ring around the central hub of the city, the Plaza de Cataluña. The area it covers is much more extensive than that of the hotels and is occupied by traditional city centre residential districts, such as El Raval, La Barceloneta, La Ribera, the Gothic Quarter, and the area around the Sagrada Familia Church. In all these zones, the presence of Airbnb is greater in relative terms than that of hotels (Figure 4c). As a result, the Airbnb accommodation contributes to increasing tourism pressure on the centre.

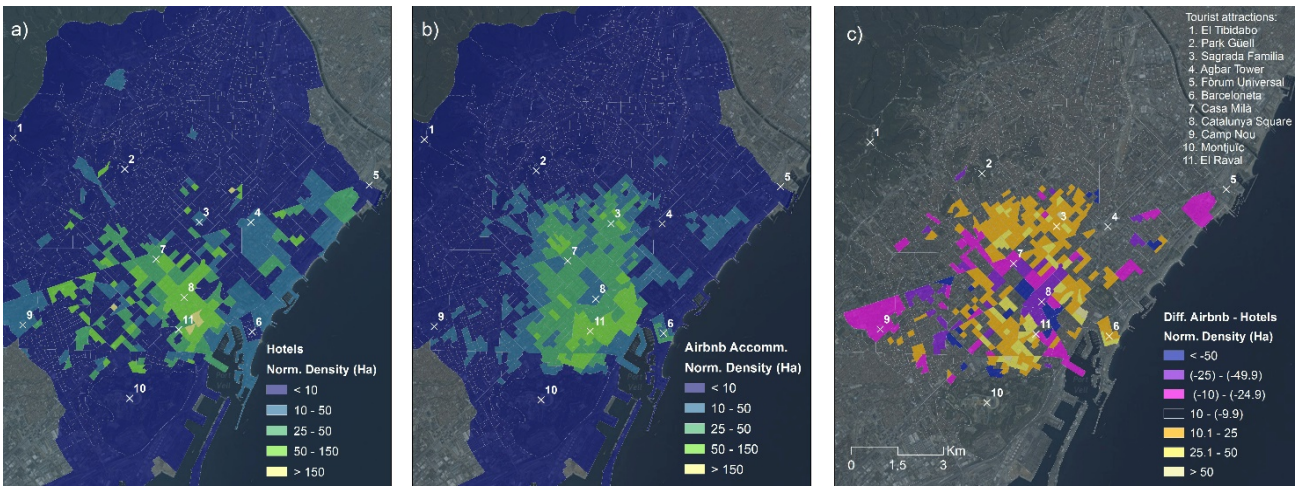

Figure 4: Density of normalised distributions: a) hotels; b) Airbnb; c) differences

Spatial statistical analysis confirms the statistical significance of these location patterns. Global Moran's Index shows a strong positive spatial correlation in both cases (positive Moran's Index and p-value = 0.00000) (Table 5), but it is higher for Airbnb than for hotels. Using Anselin Local Moran's I statistic, the spatial cluster distribution can be identified (Figure 5). As expected, both cases show a clear concentration of HH clusters in the city centre and LL on the periphery. In the case of hotel accommodation, the HH clusters are located along the main Ramblas-Paseo de Gracia axis, around which LH outliers appear. These are central areas that have traditionally had a marked residential character but no hotel accommodation. In the case of Airbnb, the HH clusters extend through all the census sections in the city centre, including those that are residential in nature. In both cases, towards the outer edge of the central area is a belt of sections with values that are not significant, which would mark the limit of tourist accommodation in the central area. In the case of Airbnb, this belt is narrow and very clearly defined and is surrounded by LL census sections in all the periphery of the municipality. In contrast, in the case of hotels, the belt with not significant values is much more diffuse and extensive because of the presence of hotels on the axis of the Diagonal and in some peripheral census sections (with not significant values or with HL outliers, which reduces the extension of LL clusters) The number of census sections according to types of cluster confirms that the distribution of Airbnb gives a greater positive spatial correlation than that of hotels, with more census sections made up of types HH and LL, and fewer outliers (LH and HL) (Table 6).

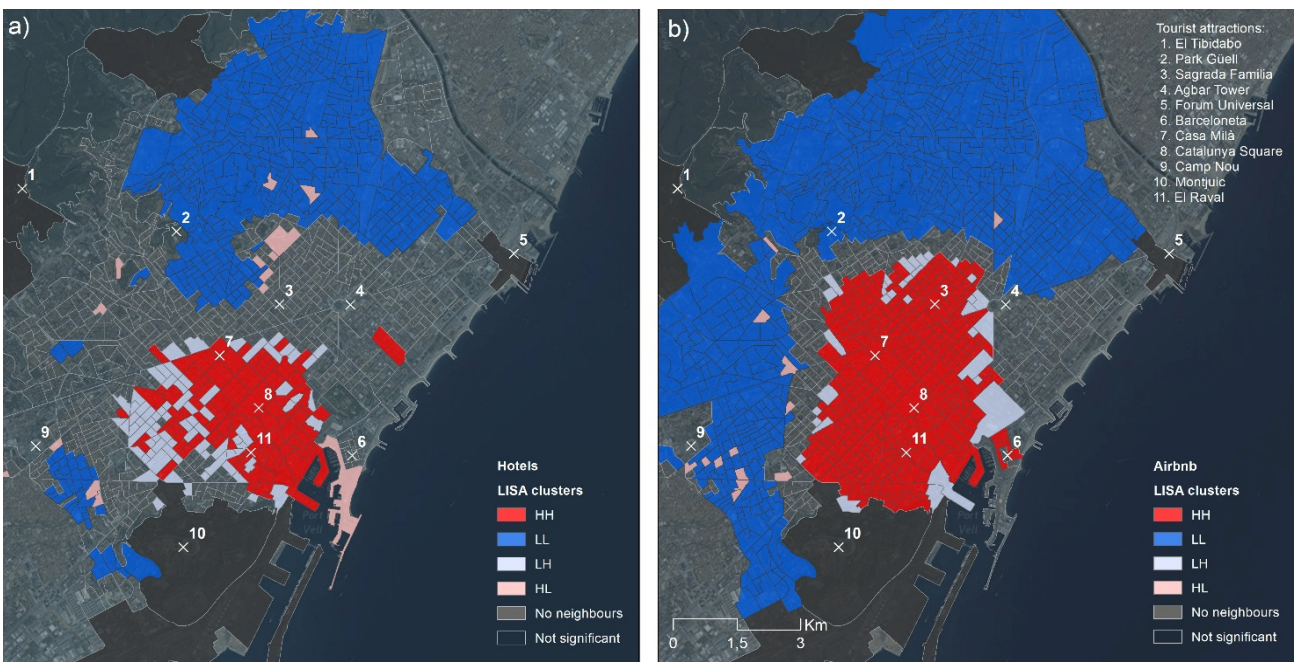

Figure 5: Anselin Local Moran's I statistic (LISA): a) accommodations in hotels; b) accommodations in Airbnb

Table 5: Global Moran's I statistics

|  | Hotels | Airbnb | Panoramio |
|---|---|---|---|
| Global Moran's Index | 0,23 | 0,70 | 0,18 |
| z-score | 27,91 | 78,55 | 25,84 |
| p-value | 0,01 | 0,01 | 0,01 |

Table 6: Number of census sections according to type of spatial cluster (LISA)

|  | Hotels | | Airbnb | | Panoramio | |
|---|---|---|---|---|---|---|
|  | Total | Percentage | Total | Percentage | Total | Percentage |
| High-High (HH) | 81 | 7.6 | 258 | 24.3 | 64 | 6.0 |
| Low-Low (LL) | 392 | 36.9 | 575 | 54.2 | 422 | 39.8 |
| Low- High (LH) | 77 | 7.3 | 33 | 3.1 | 62 | 5.8 |
| High -Low (HL) | 15 | 1.4 | 13 | 1.2 | 17 | 1.6 |
| Not Significant | 496 | 46.7 | 182 | 17.2 | 496 | 46.7 |
| Total | 1061 | 100.0 | 1061 | 100.0 | 1061 | 100.0 |

### 5.2. Areas visited by tourists

In order to analyse the main tourist areas, we mapped the density of photographs taken by tourists in terms of both absolute values (photographs/ha) (Figure 6) and normalised values (Figure 6b) to facilitate comparisons with the density of hotel and Airbnb accommodations. The photographs reflect the spatial distribution of the city's main sightseeing spots . The most photographed places, and consequently the most visited, are the Barcelona of Gaudi (the Sagrada Familia Church, Casa Batlló, Casa Milà, the Güell Park, etc.), the Gothic Quarter in the historic centre, the port area and the beach, together with other tourist spots, such as Barcelona Football Club's Nou Camp stadium, the Torre Agbar and Forum buildings, or green spaces with scenic views like Montjuic and Tibidabo. This all shows a distribution of census sections that are dispersed throughout the city and have very high intensities, in which values exceed more than 2000 and even 3000 photos, compared to an average of 25 photographs per section (Figure 6a and Table 7).

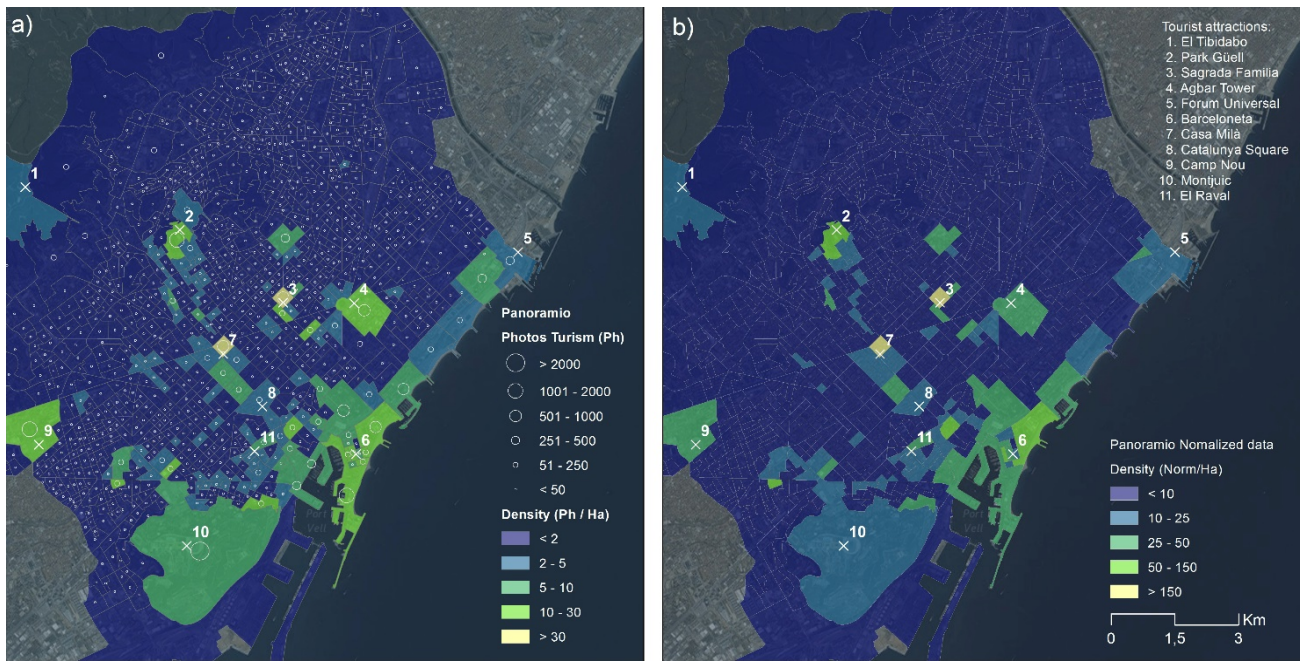

Figure 6: Photographs taken by tourists: a) total number and density; b) normalized data

Table 7: Basic statistics on the distribution of photographs taken by tourists according to census sections

|                      | Total photographs | Photographs/ha |
|----------------------|------------------:|---------------:|
| Count:               | 1061              | 1061           |
| Minimum:             | 0                 | 0              |
| Maximum:             | 3827              | 70.9           |
| Sum:                 | 26649             | 1176.8         |
| Mean:                | 25.1              | 1.1            |
| Standard Deviation:  | 161.2             | 3.3            |
| CV:                  | 641.8             | 300.8          |

Moran's Index indicates a strong positive spatial autocorrelation in the distribution of areas, although with a lower value than in the case of Airbnb and hotels (Table 5). Calculation of Anselin Local Moran's I statistic (Figure 7) shows two zones in which HH clusters are located, one in the area round Las Ramblas and the Gothic Quarter in the city's historic centre, and the other to the north of this, centred around the Sagrada Familia and including the Paseo de Gracia axis. Logically, in the more peripheral census sections away from the coast, the presence of tourists is diluted and LL clusters predominate.

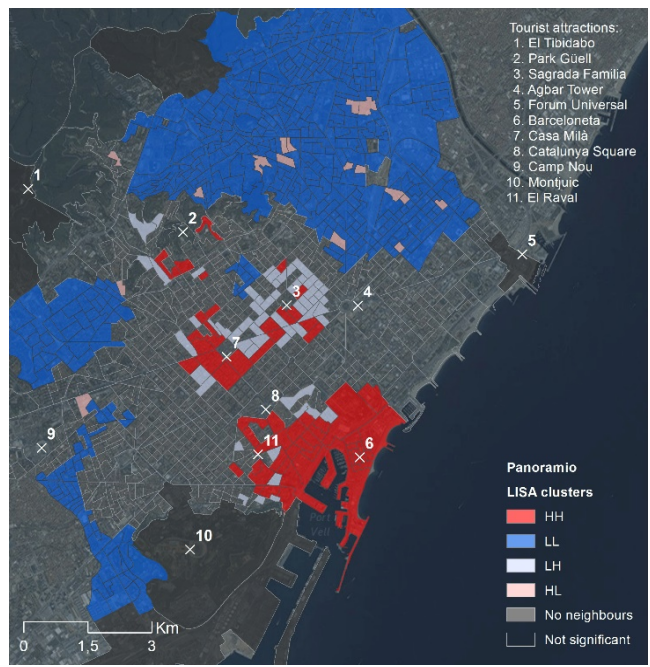

Figure 7: Anselin Local Moran's I statistic (LISA) for the distribution of photographs taken by tourists

### 5.3. Relations between the locations of different types of lodgings: hotels vs Airbnb

Relations between the location patterns of hotel and Airbnb accommodation can be analysed using bivariate autocorrelation indicators, both the Moran Index and LISA. The bivariate Moran Index shows a very high positive spatial autocorrelation between location of the hotels and accommodations offered by Airbnb (Table 8). The mapped clusters are shown in Figure 8 and the number of census sections by type are in Table 9. Type HH is located along the Ramblas-Plaza de Cataluña-Paseo de Gracia axis. These census sections have a large number of hotel places and are surrounded by sections with a high supply of Airbnb accommodation. Census sections of this type also appear in traditional residential districts in the centre, where some hotels are found in areas with a very strong Airbnb presence. Nevertheless, in these central residential districts LH census sections predominate, with a low number of hotel places and a high level of Airbnb accommodation. On the periphery, LL clusters prevail, that is, census sections with a low number of hotel places surrounded by sections with a low Airbnb presence. There are only a few cases of HL type census sections (high concentration of hotels and low presence of Airbnb), one example being the Diagonal axis.

To summarise, the distribution of clusters shows a clear centre-periphery pattern. After the centre (HH) come successive belts of LH, not significant and LL values. The high number of HH and LL clusters proves the close spatial association between the two types of accommodation offers. The distribution of outliers is very typical: LH outliers in the city centre around the big hotel axis (Airbnb expansion) and HL outliers in the LL belt (hotels dispersed on the periphery).

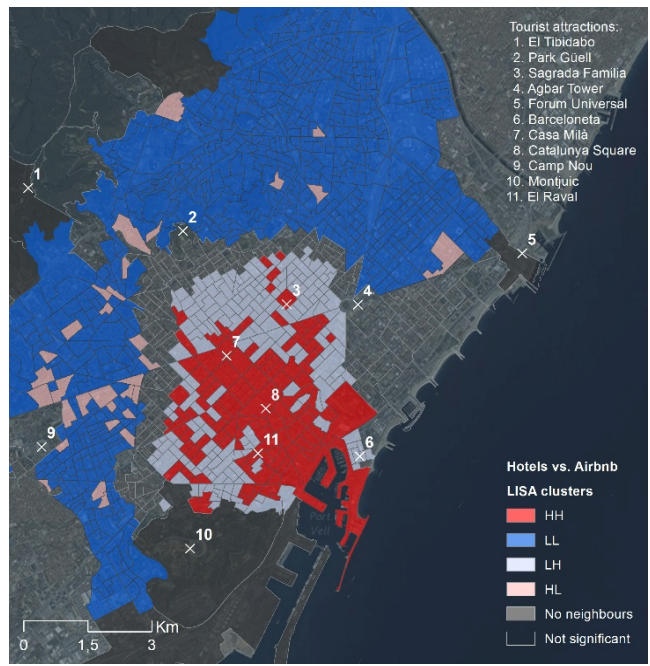

Figure 8: Bivariate Anselin Local Moran's I statistic between offers of hotel and Airbnb accommodations.

Table 8: Bivariate Global Moran's I statistics (* Significant at the 0.01 level)

|  | Hotels-Airbnb | Hotels - Panoramio | Airbnb - Panoramio |
|---|---|---|---|
| Global Moran's Index | 0.32 | 0.08 | 0.18 |
| z-score | 57.70* | 15.48* | 25.,84* |

Table 9: Number of census sections according to types of cluster with respect to the location of hotel and Airbnb

|  | Relation Hotels – Airbnb | |
|---|---|---|
|  | Total | Percentage |
| High-High (HH) | 106 | 10.0 |
| Low-Low (LL) | 557 | 52.5 |
| Low- High (LH) | 191 | 18.0 |
| High -Low (HL) | 32 | 3.0 |
| Not Significant | 175 | 16.5 |
| Total | 1061 | 100.0 |

### 5.4. Sightseeing spots and accommodation

Tourists tend to stay in places close to areas where the main sights and other tourist attractions are situated. Therefore, it is only to be expected that there is a strong spatial association between location of the accommodation (in hotels and with Airbnb) and the areas of the city that are of interest to tourists (photographs taken by tourists). Although the bivariate Moran's I confirms a very strong positive spatial autocorrelation in both cases, this is greater for Airbnb (Table 8), which suggests a better location of this type of accommodation with respect to the city's tourist attractions.

The bivariate LISA (Figure 9 and Table 10) shows that the HH census sections are more numerous in the Airbnb-Panoramio relation than in the Hotel-Panoramio relation. These are census sections with a high density of accommodations surrounded by sections with a high density of photographs. In the case of hotels, HH are in the census sections of the Ramblas-Gothic Quarter, and around the Paseo de Gracia (with tourist sites at Casa Milà-La Pedrera and Casa Batlló). With respect to Airbnb, HH are located along these same axes, but they also extend in particular to the residential district of the Ensanche around the Sagrada Familia. With respect to hotels, these same census sections form part of type HL (low number of hotels in areas with a high number of photographs). There is a marked contrast between those results obtained for hotels and those for Airbnb in the census sections comprising the port, La Barceloneta and the beach. With respect to the relation between the hotels and the photographs, the appearance of HH census sections in the port and beach areas is due to the presence in this zone of hotels that are surrounded by much-photographed sites, with the exception of the traditional La Barceloneta district, which has no hotels and appears as HL. The opposite occurs in the case of Airbnb, since the LH census sections predominate in the port and beach zones, while La Barceloneta is HH. Around these central areas are census sections with values that are not significant. LL predominate in peripheral locations towards the outer edge, particularly in the case of Airbnb, while in that of hotels there is a greater presence of HL census sections due to hotels scattered in areas where few photographs are taken.

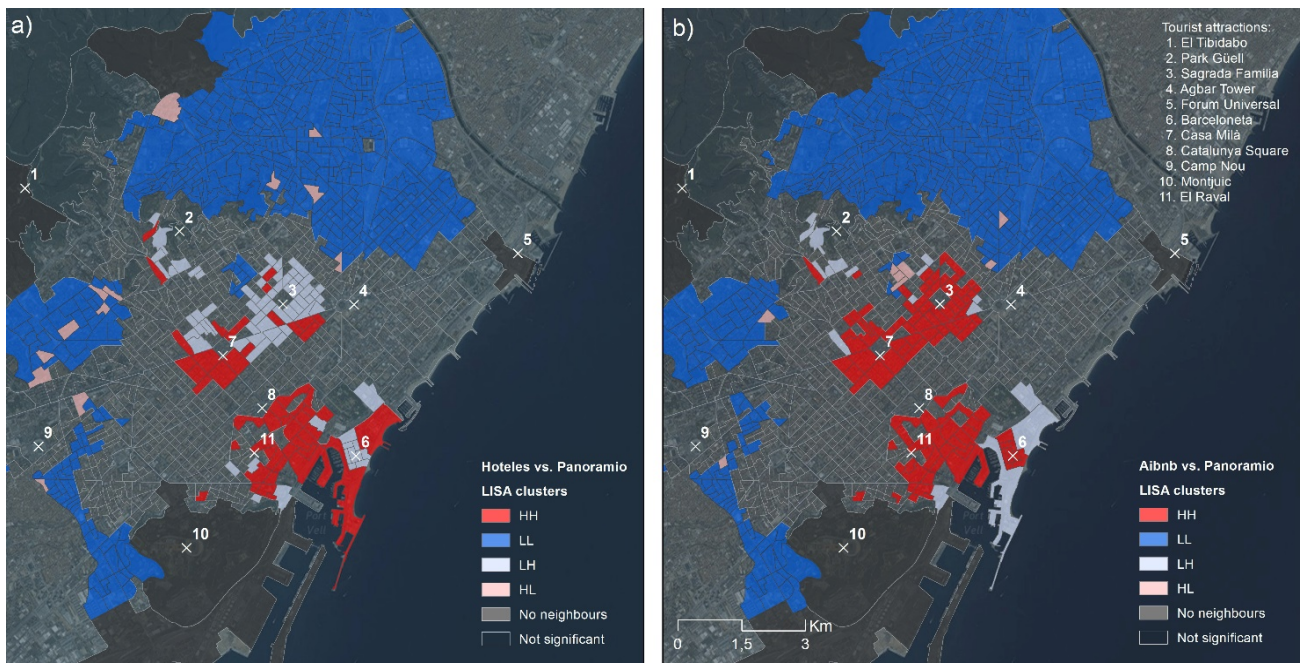

Figure 9: Bivariate Anselin Local Moran's I statistic for hotels-Panoramio and Airbnb-Panoramio

Table 10: Bivariate LISA. Number of census sections according to types of cluster in relation to hotel-Panoramio location and Airbnb-Panoramio location.

|  | Hotels- Panoramio | | Airbnb- Panoramio | |
| --- | --- | --- | --- | --- |
|  | Total | Percentage | Total | Percentage |
| High-High (HH) | 50 | 4.7 | 102 | 9.6 |
| Low-Low (LL) | 430 | 40.5 | 438 | 41.3 |
| Low- High (LH) | 78 | 7.4 | 24 | 2.3 |
| High -Low (HL) | 13 | 1.2 | 8 | 0.8 |
| Not Significant | 481 | 45.3 | 480 | 45.2 |
| Others * | 9 | 0.9 | 9 | 0.9 |
| Total | 1061 | 100.0 | 1061 | 100.0 |

**\*** Census sections larger than 2 km wide (without neighbours within 1 km from the centroid)

## 5.5. Pressure from tourism on residential areas

In order to analyse tourism pressure on residential areas, the number of places of accommodation (hotels and Airbnb) per 1,000 inhabitants has been calculated according to census sections, excluding those sections that have hardly any population (< 5 inhabitants/ha), such as green spaces or industrial zones (Figure 10).

Figure 11a highlights the tourism pressure exerted by hotels along the main hotel axis, where several census sections exceed 500 accommodations per 1000 inhabitants. Between this axis and the periphery there is an abrupt drop in pressure on residential areas from tourist accommodation. This drop is much more gradual in the case of Airbnb (Figure 11b). The lodgings available through Airbnb extend over residential areas in the centre which have no hotels and where there were formerly no accommodations. Some census sections have more than 100 Airbnb places per 1000 inhabitants, reaching a maximum of almost 400 places per 1000 inhabitants. The pressure from this type of accommodation on the centre is intensified by the fact that Airbnb occupation levels are greater in the centre than on the periphery (see Section 3). Finally, Figure 10c shows the total number of places for both types of accommodation (hotels and Airbnb). This map should be analysed with caution, since the level of occupation is higher for hotels than for Airbnb, but it illustrates the pressure exerted by tourism on the city due to Airbnb lodgings, with several census sections in which the number of places exceeds and even doubles the number of inhabitants (see also Table 11).

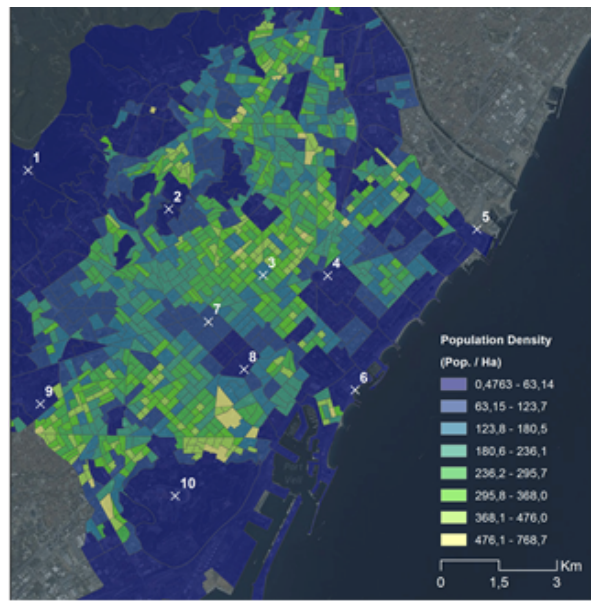

Figure 10: Population density according to census sections (inhabitants/ha).

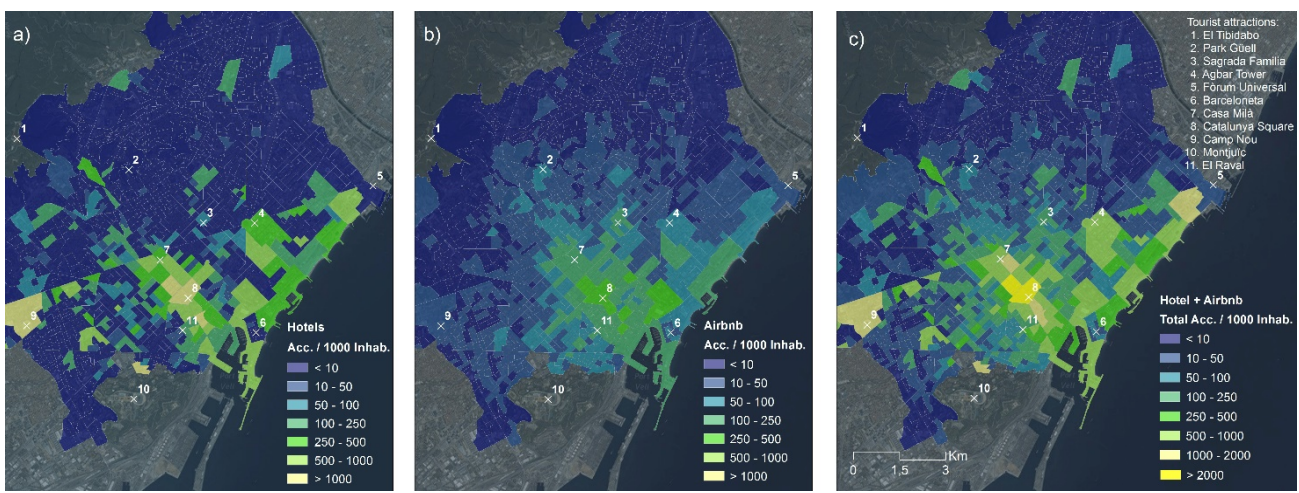

Figure 11: Tourism pressure on residential areas: a) Hotel places per 1000 inhabitants; b) Airbnb places per 1000 inhabitants; c) Hotel places + Airbnb places per 1000 inhabitants.

Table 11: Basic statistics on the distribution of the accommodations offered by hotels and Airbnb per 1000 inhabitants according to census sections.

|  | Hotel places / 1000 inhabitants | Airbnb places /1000 inhabitants | Hotel + Airbnb places / 1000 inhabitants |
|---|---:|---:|---:|
| Count | 1061 | 1061 | 1061 |
| Minimum | 0 | 0 | 0 |
| Maximum | 1796.1 | 391.7 | 2148 |
| Sum | 46079.3 | 31941.6 | 78021 |
| Mean | 43.4 | 30.1 | 73.5 |
| Standard Deviation | 150.0 | 45.9 | 180.6 |
| CV | 345.4 | 152.5 | 245.6 |

## 6. Conclusions

The irruption of P2P accommodation platforms in tourist cities has received very little attention from researchers, particularly in relation to the location of its accommodations and their possible impact on the city. The present study aims to close this gap with a contribution on the location of Airbnb accommodation offered in cities with mass tourism, related to hotel location and the most visited tourist attractions (locational advantages), and the resident population (tourism pressure), using Barcelona, a city with one of the highest numbers of tourists in Europe, as a case study.

The results of the study show that the distribution of the Airbnb accommodations offered in Barcelona has a clear centre-periphery pattern. Its listings tend to be concentrated in the city centre, where they cover a wider area than the main hotel axis as they also extend to very central residential districts that are not covered by hotel lodgings. Spatial autocorrelation analysis shows that the spatial distribution of Airbnb accommodations has a statistically significant greater positive spatial autocorrelation than that of hotels. The distribution of Airbnb is much simpler and more regular, from the HH clusters in the centre to the LL clusters at the periphery, incorporating a narrow band of not significant census sections, and with a scarcity of outliers. In contrast, hotels show more complex patterns, with less extension of HH and LL clusters and a greater extension of not significant census sections and outliers.

Bivariate spatial autocorrelation analysis reveals a close spatial association between the Airbnb accommodations offered and that of hotels. The centre-periphery patterns shown by the results of this analysis are very clear. The axis of hotels in the centre (cluster HH) gives way to a band dominated by Airbnb (LH), a second band of not significant values and finally a peripheral area dominated by LL clusters, with some HL outliers (areas with a concentration of hotels but not Airbnb). In short, the census sections with HH and LL clusters and not significant values indicate a similar behaviour for both types of accommodation in much of the city. It is in the outliers where differences are found, with central locations where Airbnb prevails (LH) and peripheral ones where there is a predominance of hotels (HL).

Analysis of the bivariate autocorrelation between the accommodations and the sightseeing spots (tourists' photographs geolocated on Panoramio) confirms a close spatial association between both variables (accommodation and places visited), with very similar distributions in the two types of accommodation. In both cases, two areas of concentration of HH clusters are identified which basically coincide with the medieval historic centre in the south and the Barcelona of Gaudi in the north. The differences between the two maps are generally to Airbnb's advantage. Airbnb has a greater number of HH census sections (high concentration of accommodation places in sections surrounded by heavily visited areas), while hotels show a greater number of outliers, revealing a worse location in relation to the most visited tourist attractions, since places are concentrated in the census sections of areas where few photographs are taken (HL) or in those with a low density of places in highly-photographed surroundings (LH). The first occurs particularly in hotels on the periphery and can be explained by the fact that these hotels are generally oriented towards business travel and therefore have other locational requirements. The second reveals the existence of census sections in which the hotel sector does not take sufficient advantage of proximity to the places most visited. In short, the results above suggest that Airbnb benefits in greater measure than hotels from proximity to the most visited places in the city (greater number of HH clusters and fewer HL outliers), probably because of its greater facility for expansion in already built-up areas.

Finally, the relation between accommodation places and the resident population shows that new residential areas are being added to the traditional areas of strong pressure from tourism along the city's main tourist axis, and Airbnb clearly contributes to that pressure. It is in these census sections where problems have arisen, involving the coexistence of the new Airbnb lodgings and the resident population, particularly in certain census sections of La Barceloneta, El Raval, the Gothic Quarter and La Ribera. The reason for these conflicts is that the expansion of Airbnb has led to ordinary rental flats being removed from the market, resulting in increased rents and processes that drive out the local population (more than half the Airbnb lodgings in Barcelona consists of entire homes/apartments). Airbnb is also transforming the business structure of these areas, as in the case of shops and restaurants, which are increasingly geared to tourists.

Airbnb is changing the tourist accommodation model in a way that, currently, creates conflict in cities with mass tourism. Barcelona is trying to control expansion of this type of rental through inspections to ensure that apartments are not functioning illegally and that taxes are paid, with fines of up to 90,000 euros imposed. In this way, not only are more taxes collected but Airbnb's competitive advantage over traditional accommodation is reduced, thereby reducing its prospects for further expansion.

## Acknowledgments

The authors gratefully acknowledge funding from the ICT Theme of the European Union's Seventh Framework Program (INSIGHT project - Innovative Policy Modeling and Governance Tools for Sustainable Post-Crisis Urban Development, GA 611307), from the Madrid Regional Government (S2015/HUM-3427) and a post-doctoral fellowship from Ministerio de Economía y Competitividad of Spain (FPDI 2013/17001).